\documentclass[amsmath,amssymb,prl,aps,twocolumn,mathbbm,superscriptaddress]{revtex4}
\usepackage{amssymb}
\usepackage{graphicx}
\usepackage{times}
\usepackage{subfigure}
\newcommand{\bra}[1]{\langle #1|}
\newcommand{\ket}[1]{|#1\rangle}

\newcommand{\para}[1]{\left( #1 \right)}

\begin{document}
\title{Relativistic Motion with Superconducting Qubits}
\author{S. Felicetti}
\address{Department of Physical Chemistry, University of the Basque Country UPV/EHU, Apartado 644, E-48080 Bilbao, Spain}
\author{C. Sab{\'\i}n}
\address{School of Mathematical Sciences, University of Nottingham, University Park NG7 2RD Nottingham, United Kingdom}
\author{I. Fuentes}
\address{School of Mathematical Sciences, University of Nottingham, University Park NG7 2RD Nottingham, United Kingdom}
\address{University of Vienna, Faculty of Physics, Boltzmanngasse 5, 1090 Wien, Austria}
\author{L. Lamata}
\address{Department of Physical Chemistry, University of the Basque Country UPV/EHU, Apartado 644, E-48080 Bilbao, Spain}
\author{G. Romero}
\address{Departamento de F\'isica, Universidad de Santiago de Chile (USACH), Avenida Ecuador 3493, 917-0124, Santiago, Chile  }
\author{E. Solano}
\address{Department of Physical Chemistry, University of the Basque Country UPV/EHU, Apartado 644, E-48080 Bilbao, Spain}
\address{IKERBASQUE, Basque Foundation for Science, Maria Diaz de Haro 3, 48013 Bilbao, Spain}

\date{\today}

\begin{abstract}
We show how the dynamical modulation of the qubit-field coupling strength in a circuit quantum electrodynamics architecture mimics the motion of the qubit at relativistic speeds. This allows us to propose a realistic experiment to detect microwave photons coming from simulated acceleration radiation. Moreover, by combining this technique with the dynamical Casimir physics, we enhance the toolbox for studying relativistic phenomena in quantum field theory with superconducting circuits.
\end{abstract}


\maketitle

Circuit quantum electrodynamics (cQED) \cite{Wallraff2004} has swiftly become one of the leading quantum technologies for testing fundamentals of quantum mechanics, and for the implementation of quantum information tasks.  The possibility of fast tuning characteristic parameters of the interaction between superconducting qubits and the electromagnetic field allows for the experimental investigation of quantum optical systems in previously inaccessible regimes, thus enabling the exploration of a wide variety of new physical phenomena \cite{yanamura, ultrastrong,casimirwilson}. In particular, cQED is a natural arena to implement quantum field theory (QFT) concepts~\cite{reviewjohansson}. Indeed, the physics associated to the dynamical Casimir effect (DCE) \textemdash the generation of photons out of the quantum vacuum through the motion of boundary conditions\textemdash has been implemented in a superconducting circuit platform. This experiment opened a new avenue of research on the analysis of the role of relativistic motion in quantum information setups \cite{rqt,Felicetti2014, twinpa, Andersen2015}. In this sense, quantum technologies are rapidly progressing from tabletop experiments to long-range networks \cite{zeilingerteleport} and may enter, in the near future, into the space-based realm \cite{spaceexp1}. In this scenario, relativistic effects are expected to become relevant~\cite{ivyalsingreview}.  

The Unruh effect~\cite{Milonni94}, a paradigmatic quantum field theory prediction, consists in the excitation of ground state atoms due to their acceleration through vacuum, followed by spontaneous emission. Such a phenomenon has never been observed, requiring unachievable accelerations in order to generate detectable signals. This effect can be enhanced by many orders of magnitude when two-level quantum systems are accelerated through a single-mode high-Q cavity~\cite{scully}.  The relation between this cavity-enhanced acceleration radiation and the original formulation of the Unruh effect in free space has produced a thought-provoking debate \cite{scullycomment, hureply, scullytheory}.

\begin{figure}[b]
\centering
\includegraphics[angle=0, width=0.45\textwidth]{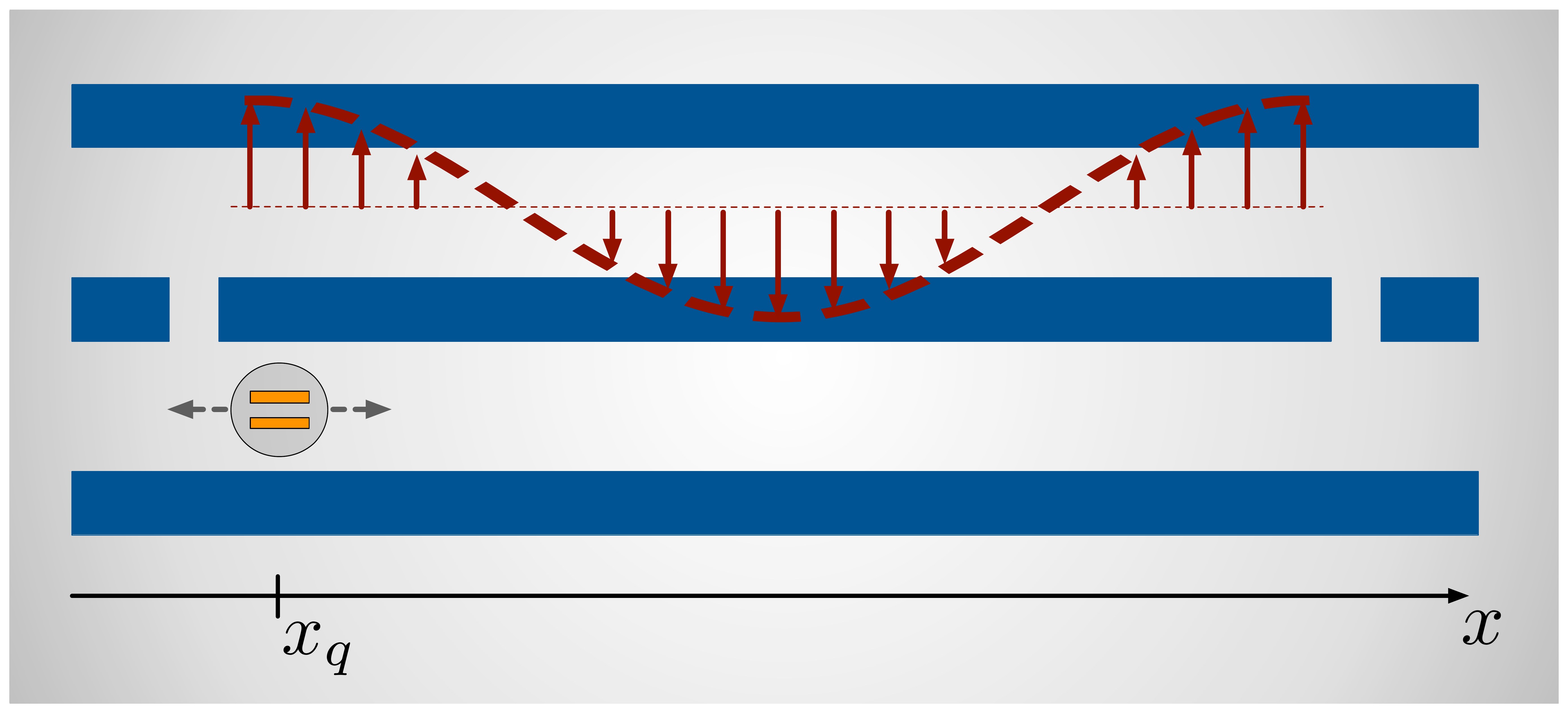}
\caption{\label{flux} (Color online) Sketch of a transmission line resonator coupled to a superconducting qubit, where the coupling strength depends on the position of the latter. The red dashed line follows the voltage profile of the second resonant mode of a $\lambda/2$ resonator.}
\end{figure}

In this Letter, we show how the ultrafast variation of the coupling strength between a superconducting qubit and a transmission line resonator mimics the motion of the qubit at relativistic speeds. This analogy paves the way for novel experimental implementations of QFT effects associated to relativistic motion. Here, we propose the measurement of the radiation produced by an artificial atom performing simulated acceleration through a superconducting cavity.  Our technique provides a novel tool for the quantum simulation of QFT in curved space times with current circuit QED technology, in addition to the observation of the DCE \cite{casimirwilson} and the possibility of generating effective spacetime metrics \cite{reviewjohansson,casimirsorin}.

We consider the system depicted in Fig.~\ref{flux}, consisting of a superconducting qubit interacting with the electromagnetic field confined within a transmission line resonator (TLR). This model can be described by the Hamiltonian
\begin{equation}
\label{physHam}
\mathcal{H} = \omega a^\dagger a + \frac{\omega_q}{2}\sigma_z + \mathcal{H}_I(x_q) ,
\end{equation}
where $\omega_q$ is the qubit energy spacing, $\sigma_z$ and $\sigma_x$ are the usual Pauli operators acting on the qubit Hilbert space, and we consider $\hbar = 1$. We assume that the system dynamics effectively involves a single resonator mode, described by annihilation and  creation operators $a$ and $a^\dagger$, respectively, of frequency $\omega= v k$ and wave vectors $k = \pi/L$. Here, $L$ is the resonator length and  $v$ is the group velocity, which is established by the resonator electrical properties. 
The interaction Hamiltonian between superconducting qubits and a transmission line resonator (TLR) is mediated either by the electrical potential $V$, for  capacitive coupling like in the case of transmon qubits, or by the current $I$ for inductive coupling in the case of flux qubits.
For the resonator stationary modes, both physical quantities have a periodic space modulation. Without loss of generality, we can write the interaction Hamiltonian as
\begin{equation}
\label{physHI}
\mathcal{H}_I(x_q) =  g \cos{\left( k x_q \right)} \sigma_x (a^\dagger + a),
\end{equation}
where $g$ is the coupling strength and $x_q$ the qubit position~\cite{Shanks2013}. We point out that in circuit QED designs, it is possible to modulate $g$ in time, by means of a controllable magnetic flux $\phi_x = f\ \phi_0$ flowing through the loop of a superconducting quantum interference device (SQUID). We introduced the frustration parameter $f$ and the magnetic flux quantum $\phi_0$. Due to the non-linear inductance of the Josephson junctions, such tunable coupling has usually a periodic dependence on~$f$, $g=g_0 \cos{(f)}$. We consider now a fixed qubit position, chosen in order to maximise the coupling strength. This is done by placing the qubit position in a node(antinode) of the current $I$ for a charge(flux) qubit, corresponding to $x_q=0$ in Eq.~\eqref{physHam}. 
Accordingly, the interaction Hamiltonian has the following dependence on $f$,
\begin{equation}\label{physHI2}
\mathcal{H}_I(f) =  g_0 \cos{\left(  f \right)} \sigma_x (a^\dagger + a).
\end{equation}
Choosing the profile for the magnetic flux
\begin{equation}\label{eq:fprof}
f= k\,x_q,
\end{equation}
the Hamiltonians in Eqs.~\eqref{physHI} and~\eqref{physHI2} are equivalent. Therefore, the modulation of the effective coupling constant mimics the motion of the qubit $x_q(t)$ inside the TLR. Later on, we will detail how this effective motion can produce qubit and field excitations in the relativistic regime. In the following, we show that fast-tuning of the coupling strength can be achieved in the framework of superconducting circuits.

\begin{figure}[]
\centering
\includegraphics[angle=0, width=0.48\textwidth]{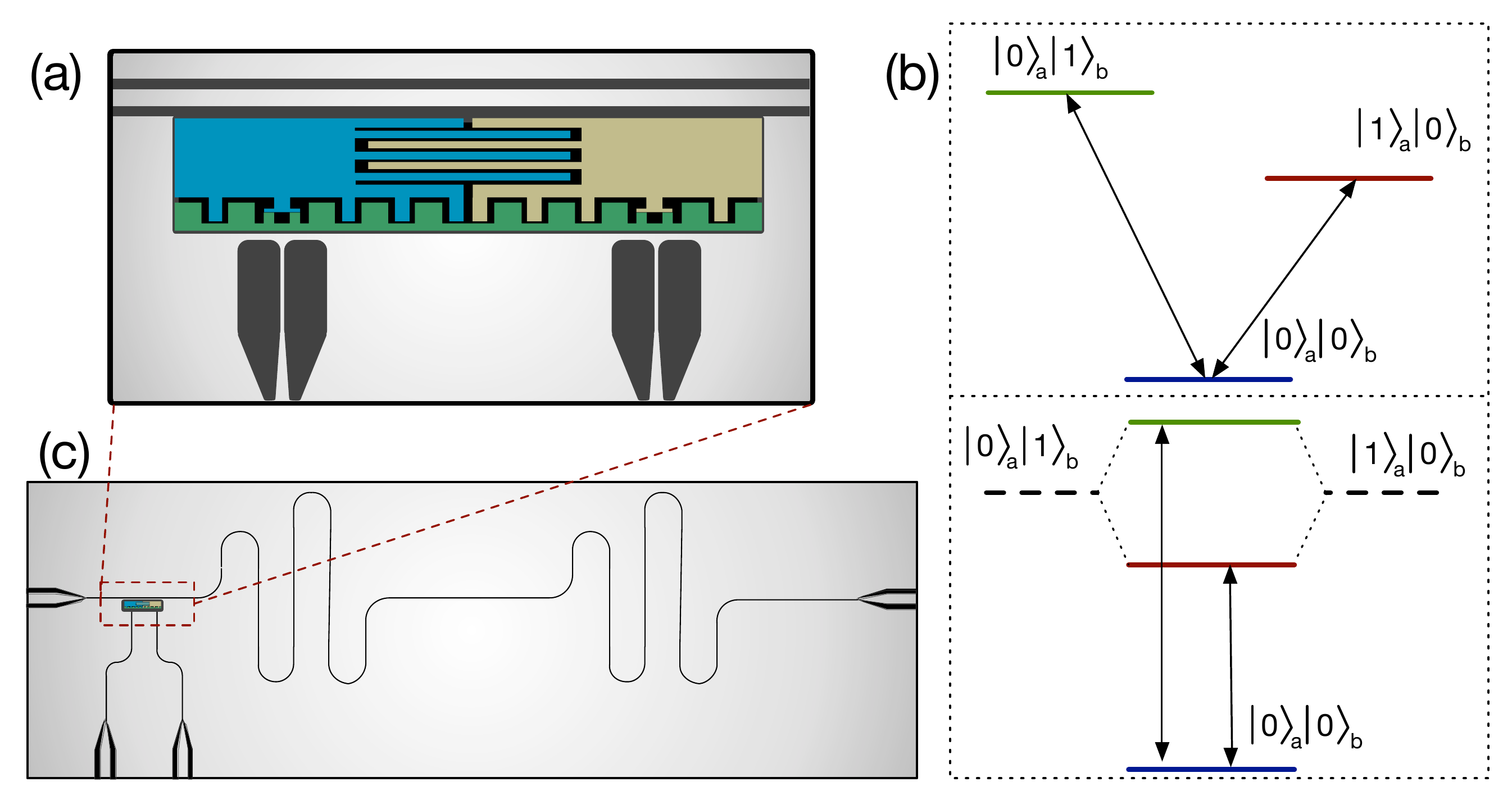}
\caption{\label{trans} 
(Color online) (a) Sketch of a modified transmon qubit, see Ref.~\cite{Srinivasan2011}. Different colors identify independent superconducting islands capacitively coupled between each other. (b) Scheme of the energy levels. The qubit is composed of two coupled transmons, the logical states are given by the two lower collective eigenstates (blue and red). The third collective energy level (green) can be used for reading the qubit state. (c) Sketch of a superconducting chip containing a TLR interacting with a  tunable coupling qubit.}
\end{figure}

There are different ways of implementing a qubit-resonator system interacting through the tunable Hamiltonian of Eq.~\eqref{physHI2}, using either phase \cite{Allman2010}, flux \cite{Romero2012,ultrastrongperop}  or transmons~\cite{Gambetta2011,Koch2007} qubits. The configuration here considered, shown in Fig.~\ref{trans}, allows independent tuning of the qubit-field interaction $g$ and of the qubit energy spacing $\omega$. Hence, it provides a versatile solution for simulating a broad range of parameter regimes with a single sample.
As shown in Fig.~\ref{trans}(a), the present qubit consists of two capacitively shunted SQUID loops, which form three superconducting islands. This design is well described with a model composed of two transmons, i.e. anharmonic oscillators, capacitively coupled with the TLR and between each other. The qubit will be formed by the two lowest collective eigenstates of such system, here labeled with $\ket{E_0}$ and $\ket{E_1}$. As shown in Fig.~\ref{trans}(b), when the two anharmonic oscillators are detuned, the qubit logical states are given by the vacuum $\ket{0}_L=\ket{E_0}=\ket{0}_a\ket{0}_b$ and $\ket{1}_L=\ket{E_1}=\ket{1}_a\ket{0}_b$, where, without loss of generality, we assume that $\omega_a<\omega_b$.  We denote with $\ket{0}_i$ and $\ket{1}_i$ the ground and first excited eigenstate of the $i$-th transmon, and with $\omega_i$ the energy spacing between such states. On the other hand, when the two anharmonic oscillators are nearly degenerate, the first collective excited level will be given by $\ket{1}_L= \ket{E_1}=\left( \ket{1}_a\ket{0}_b - \ket{0}_a\ket{1}_b \right)/\sqrt{2}$. Such a state has no dipole coupling with the TLR, so in the latter configuration the cavity-qubit interaction is strongly inhibited~\cite{Gambetta2011}. 

Controlling the external magnetic flux flowing through the two SQUID loops allows to independently tune the level spacing of the two transmons. This enables us to swap between the two collective energy configurations and, at the same time, to control the logical qubit transition frequency.
Using this configuration, tuning of the cavity-qubit coupling strength from hundreds of KHz to tens of MHz has been experimentally proven \cite{Srinivasan2011}.
We also point out that, in the nearly degenerate case ($\omega_a \approx \omega_b$), the second excited collective level is given by $\ket{2}=\ket{E_2}=\left( \ket{1}_a\ket{0}_b + \ket{0}_a\ket{1}_b \right)/\sqrt{2}$. This state experiences a strong dipole coupling with the cavity mode, hence it can be used for fast-reading the qubit state. Such a measurement scheme consists in driving the transition $\ket{E_1}\rightarrow \ket{E_2}$ and then measuring the shift of the cavity resonant frequency due to the qubit state. Notice that the transition to the third level is safely detuned by at least 1 GHz from the primary qubit transition \cite{Srinivasan2011}.

On-the-fly tuning of the coupling strength over nanosecond time scales has already been demonstrated in circuit QED architectures \cite{Bialczak2011}.
Fast modulation of the cavity-qubit coupling strength has also been proposed for similar designs~\cite{Mezzacapo2014}.
In our scheme, the effective simulation of the qubit position is achieved through the modulation of the magnetic flux threading SQUIDs, that is, similarly to the motion of the boundary conditions in the DCE experiment \cite{casimirwilson}. This allows to achieve the same regime of  velocities and accelerations, for the mirrors as well as for the qubits, and it enables the possibility of joint qubit-cavity motion. Using realistic parameters, we can consider trajectories of constant simulated  acceleration ${\mathcal A} =10^{15}\,m/s^2$ during $1\,\operatorname{ns}$  \cite{twinpa},  or harmonic motion with maximum accelerations of ${\mathcal A} =10^{17}\,\operatorname{m/s^2}$ and maximum velocities of around $v/4$, where $v=1.2\times10^8\,m/s$ is the group velocity of the electromagnetic field.

We propose now the detection of simulated cavity-enhanced acceleration radiation. In this sense, a two-level atom in its ground state accelerated along a single-mode cavity has a certain probability of emitting a photon and jump to its excited state, even if the cavity is in the vacuum state~{\cite{scully}. This is due to the fact that the acceleration activates the counterrotating terms of the Hamiltonian. The cavity acceleration radiation is characterised by the absorption and emission coefficients $R_{\mathrm abs,em}$, which are related with the transition amplitudes between $\ket{g\,0}$ (qubit in the ground state, field in the vacuum) and $\ket{e\,1}$, (qubit excited, one photon), $R_{\mathrm em}=|\int^{t_1}_{t_0}dt\bra{e1}\mathcal{H}_I(x_q,t)\ket{g0}|^2$ and between $\ket{g\,2}$ and $\ket{e\,1}$, $R_{\mathrm abs}=|\int^{t_1}_{t_0}dt\bra{e1}\mathcal{H}_I(x_q,t)\ket{g2}|^2$, where we write $\mathcal{H}_I$ in the interaction picture. 
Up to second order in perturbation theory, we can relate these magnitudes with observables such as  the qubit excitations $\langle\sigma_z\rangle=1-\,R_{\mathrm em}$ and the average number of photons $\langle n\rangle_{ph}=R_{\mathrm em}(1+2\,R_{\mathrm abs}+R_{\mathrm em})$.

\begin{figure}[t]
\centering
\includegraphics[angle=0, width=0.48\textwidth]{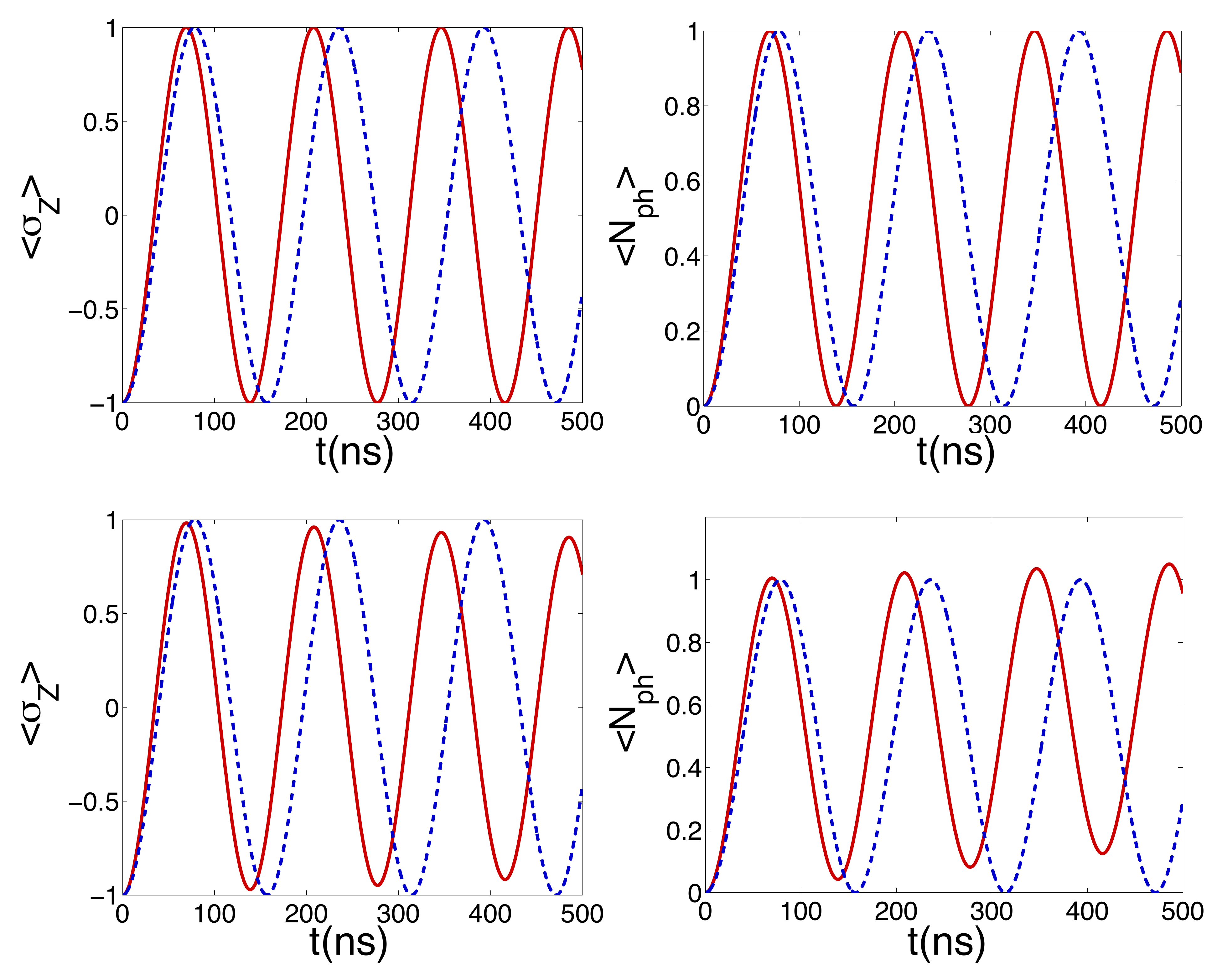}
\caption{\label{comparison}(Color online) Expected value of $\sigma_z$ and average photon number over evolution time, for a numerical simulation of the system dynamics (red continuous line) and for the exact anti-JC model (blue dashed line). We consider a resonant qubit ($\omega=\omega_q=2\pi \times 4\,\text{GHz}$), the qubit-cavity coupling constant is given by
$g(t) = g_0\, \cos[\pi/2 + \pi/2 \ \cos{\para{ \omega_d t  }  }]$, with  $g_0/\omega = 0.01$, and $\omega_d= 2\omega$. 
The upper plots show unitary dynamics, while the lower ones include dissipative qubit decay characterized by $\gamma/2\pi=400\,\text{KHz}$.}
\end{figure}

In the case of a uniformly accelerated atom crossing the cavity in the longitudinal direction, the relativistic description of the atomic qubit trajectory is given by
\begin{equation}\label{eq:position}
x_q(t)=\frac{c}{{\mathcal A}}\sqrt{c^2+{\mathcal A}^2\,t^2}.
\end{equation}
Accordingly, from Eq.~\eqref{physHI} we  have that
\begin{equation}
R_{\mathrm em}=  \left| \int^{t_1}_{t_0}dt\, g \cos{[k\,x_q(t)]}e^{-i\,\omega_q\,t}e^{-i\,\omega\,t}e^{-\gamma\,t}\right|^2,
\label{eq:Re}
\end{equation}
where we introduced the atomic qubit spontaneous decay rate $\gamma$. Notice that the observation of such a phenomenon with current quantum optical tools is unachievable in the lab, due to the extreme acceleration needed in order to produce a detectable signal~\cite{scully}. 

Let us then examine specific physical phenomena that could be simulated with the circuit QED architecture of Fig.~\ref{trans}. For uniform accelerations, from Eqs.~\eqref{eq:fprof} and~\eqref{eq:position}, we have $\frac{f}{k} = \frac{c}{{\mathcal A}}\sqrt{c^2+ {\mathcal A}^2\,t^2}$. For a resonant qubit, given the cavity frequency $\omega_q=\omega= 2\pi\times 4\ \text{GHz}$, the maximum achievable coupling strength and acceleration are $g_0/\omega = 0.01$ and ${\mathcal A}=10^{15}\,m/s^2$, respectively. Accordingly, from Eq.~\eqref{eq:Re}, the probability of qubit excitation due to relativistic motion is of the order of $R_{\mathrm em} \approx 10^{-4}$ after one trip. In order to generate detectable excitations with realistic accelerations and coupling strengths, the qubit flying time into the cavity must then be increased. Hence, a natural possibility is to consider a qubit trajectory given by harmonic oscillations. From the Hamiltonian of Eq.~\eqref{physHI2} and the relation in Eq.~\eqref{eq:fprof}, we note that such dynamics can be simulated taking the external flux $f$ to be oscillating with frequency $\omega_d$ and amplitude $\Delta f$, around a fixed offset $f_0$,
\begin{equation}
\label{geff}
f(t) = f_0 + \Delta f \cos{\para{\omega_d t}} =k\,x_q(t).
\end{equation}
In this case, the absorption and emission coefficient read
\begin{equation}
R_{\mathrm em, abs}= \left|\int^{t_1}_{t_0}dt\,g(t)\,e^{-i\,\omega_q\,t}e^{\mp i\,\omega\,t}e^{-\gamma t}\right|^2,
\end{equation} 
where $g(t) = g_0 \cos{[k\,(x_0+\Delta x \,\cos{\omega_d\,t})]}$.
Here, $x_0=f_0/k$ is the equilibrium position, while $\Delta x=\Delta f/k$ and $\omega_d$ are the amplitude and the frequency of the oscillations, respectively.

Notice that, when $g(t)$ follows harmonic oscillations at the sum of the frequencies of the cavity and the qubit, the counter-rotating transitions $\ket{g\,0}\rightarrow\ket{e\,1}$, $\ket{e\,1}\rightarrow\ket{g\,0}$ become resonant, and the anti-Jaynes Cummings (anti-JC) model is obtained. This is the case of a qubit crossing the cavity with no acceleration and at the double of the speed of light in the material. This condition makes such dynamics not implementable with current superconducting technology. However, if we set $f_0 = \pi/2$,  $\Delta f = \pi/2$ and $\omega_d = 2 \omega$,  the system behaves as the anti-JC model with effective coupling strength slightly larger than $g_0$. This can be understood by developing the coupling time-dependence in a power series, where only frequency components rotating at $~2\omega$ have  non-negligible effect on the system dynamics.The effective motion of the qubit, when coupled to the second resonant mode of a $\lambda/2$ resonator,  is given by [see Eq.~\eqref{eq:fprof}] $x_q(t)=L/2\,\left[1+\,\cos(\omega_d\,t\right)]$. Hence, such a simulation corresponds to a qubit oscillating between the center of the resonator and one of the two TLR capacitors.

In Fig.~\ref{comparison}, it is shown a comparison between  numerical simulations of such dynamics and the exact anti-JC model. The initial state for the simulation is the vacuum, i.e., the qubit in its ground state and no photon in the resonator. Regardless of the similarities between these models, we have found a parameter regime showing single-excitation transfer between the cavity and the qubit. Interestingly, by introducing dissipation in the system one can force the anti-JC dynamics out of the single-excitation subspace, as shown in Fig.~\ref{comparison}. This results in photon growth induced by the joint effect of counterrotating terms and dissipation, as studied in Ref.~\cite{Werlang2008}. 

On the other hand, if the frequency of the oscillations matches half the sum of the cavity and qubit frequencies, the counterrotating transitions are no longer resonant and experience similar behavior as the rotating ones. Therefore, the transition $\ket{e\,1}\rightarrow\ket{g\,0}$   has a similar probability as $\ket{e\,1}\rightarrow\ket{g\,2}$, resulting in a parametric photon creation and amplification.  In Fig.~\ref{comparison2}, we observe this phenomenon when $f_0=\pi$, $\Delta f = \pi$ and $\omega_d = \omega=\omega_q$. Such configuration corresponds to a qubit bouncing back and forth between the two ends of a $\lambda/2$ resonator. This photon generation can be related to the DCE \cite{dodonovdce}, this time related to the relativistic motion of the qubit.

\begin{figure}[t]
\centering
\includegraphics[angle=0, width=0.48\textwidth]{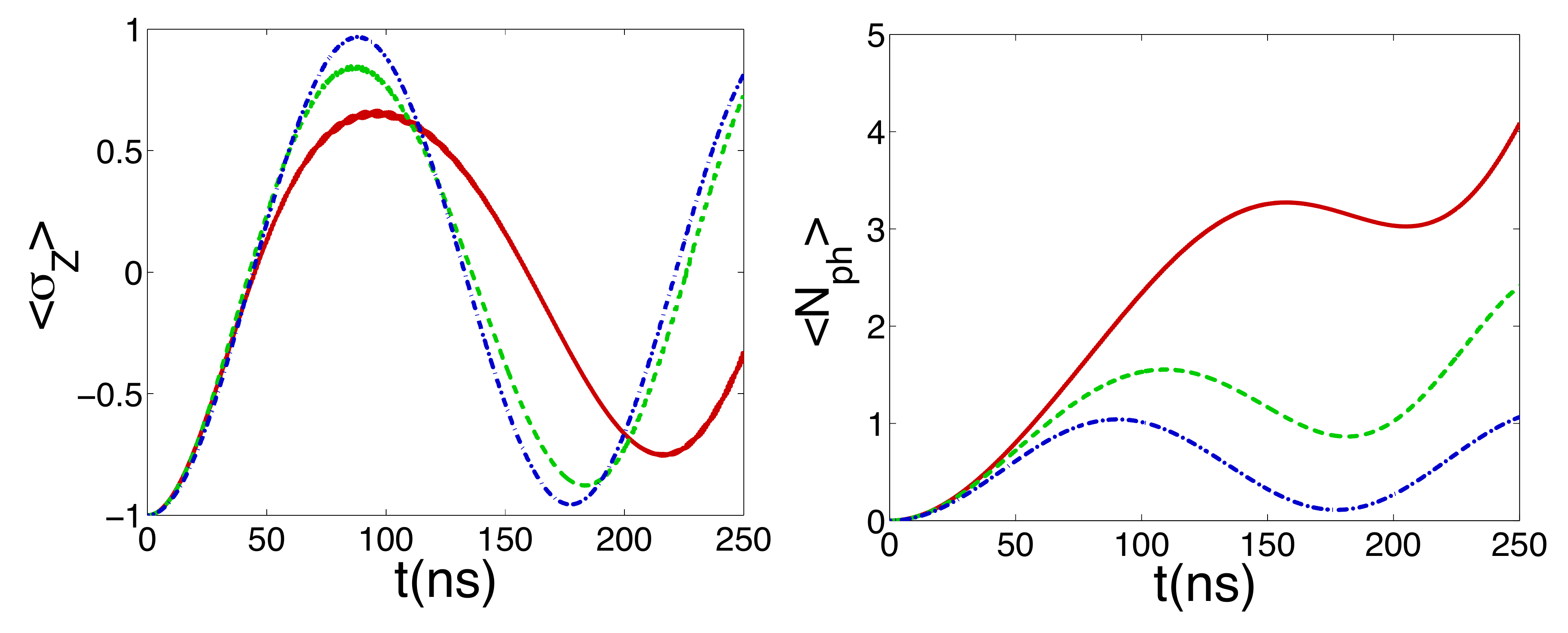}
\caption{\label{comparison2} (Color online) Numerical simulations for a resonant qubit ($\omega_q= \omega =2\pi\times 4\,\text{GHz}$) and decay rate $\gamma/2\pi= 400\,\text{KHz}$. The coupling strength is given by $g(t) = g_0\cos[ \pi + \Delta f  \ \cos{\para{ \omega_d t  }  }]$, with $g_0/\omega=0.01$. Here, $\Delta f = \pi$ (red continuous line), $\Delta f =0.9 \pi$ (green dashed line) and $\Delta f = 0.8\pi$ (blue dashed-dotted line). The value $\Delta f= \pi$ corresponds to a qubit bouncing back and forth, spanning all the cavity length.}
\end{figure}

Finally, we point out that the proposed model could be implemented with other quantum technologies that display high-frequency tunability properties. For example, the sharp frequency dependence of giant artificial atoms~\cite{Kockum2014} may be also used to tune the effective coupling strength with negligible effect on the qubit frequency. Using hybrid technologies~\cite{Gustafsson2014}, the difference in frequency scales could be exploited favourably to increase the ratio between the tuning frequency and the system characteristic frequencies.

In summary, we have shown that ultrafast modulation of the coupling strength between a superconducting qubit and a single mode of a superconducting resonator mimics the effective motion of the qubit at relativistic speeds. When the qubit follows an effective oscillatory motion, we find two different regimes. Depending on the oscillation frequency the system resembles the anti-JC dynamics or experiences unbounded photon generation. Our proposal provides an essential tool for the quantum simulation of relativistic QFT and, thus, the interplay between quantum physics and relativity.

\begin{acknowledgments}
This work was supported by the Spanish MINECO FIS2012-36673-C03-02; Ram\'on y Cajal Grant RYC-2012-11391, UPV/EHU EHUA14/04, UPV/EHU UFI 11/55; Basque Government IT472-10; CCQED, PROMISCE, and SCALEQIT European projects; and Fondecyt 1150653.
\end{acknowledgments}

\end{document}